# SEMICONDUCTOR CHARACTERIZATION BY SCANNING ION BEAM INDUCED CHARGE (IBIC) MICROSCOPY.

E. Vittone[1,2,*], Z. Pastuovic[3], P. Olivero[2], C. Manfredotti[1,2], M. Jaksic[3], A. Lo Giudice[2], F. Fizzotti[2], E. Colombo[1,2].

[1] INFN sezione di Torino, via P.Giuria 1, 10125 Torino, Italy.

[2] Dipartimento di Fisica Sperimentale and Centro di Eccellenza NIS, Università di Torino, via P. Giuria 1, 10125 Torino, Italy.

[3] Department of Experimental Physics, Ruđer Bošković Institute, P.O. Box 180, 10002 Zagreb, Croatia

**Abstract**

The acronym IBIC (Ion Beam Induced Charge) was coined in early 1990's to indicate a scanning microscopy technique which uses MeV ion beams as probes to image the basic electronic properties of semiconductor materials and devices. Since then, IBIC has become a widespread analytical technique to characterize materials for electronics or for radiation detection, as testified by more than 200 papers published so far in peer-reviewed journals. Its success stems from the valuable information IBIC can provide on charge transport phenomena occurring in finished devices, not easily obtainable by other analytical techniques. However, IBIC analysis requires a robust theoretical background to correctly interpret experimental data. In order to illustrate the importance of using a rigorous mathematical formalism, we present in this paper a benchmark IBIC experiment aimed to test the validity of the interpretative model based on the Gunn's theorem

[*] Corresponding author: e-mail: vittone@to.infn.it, Phone: +00 39 0116707371, Fax: +00 39 0116691104

and to provide an example of the analytical capability of IBIC to characterize semiconductor devices.

**Classification codes and keywords**

**Introduction**

The measurement of the charge induced at a sensing electrode by the motion of free carriers generated in a semiconductor material by MeV ion beams can be dated back to the 1930's [1], when the first studies on radiation induced conductivity were carried out on AgCl crystals. Since then, the analysis of the charge pulses generated by MeV ions (mainly alpha particles from a radioactive source), greatly contributed to determine the value of the fundamental electronic features of silicon and wide band gap semiconductors. The availability of MeV ion accelerators, which can provide more controlled beams, different ion species and energies, triggered a renewed interest in this technique, mainly motivated by the possibility of probing samples at different depths. With the advent of highly focused ion beams, IBIC spectroscopy turned into a spectro-microscopic technique able not only to measure, but also to image basic transport properties, dislocations, defects and buried junctions in semiconductors materials and devices. Comprehensive reviews of the main IBIC applications can be found in the fundamental book [2] and in a more updated publication [3].

Although the theory underlying the formation of IBIC signal was developed more than 70 years ago [4],[5], it is not rare to find in literature improper or incomplete interpretations of the charge collection mechanism, which often prevents the full exploitation of the analytical capability of IBIC.

This paper illustrates a benchmark IBIC experiment aimed to validate the mathematical model based on the Gunn's theorem and to provide evidence of the usefulness of the relevant algorithms for a comprehensive analysis and evaluation of basic electronic properties of semiconductors.

After an introduction to the charge induction mechanism, the basic differential equations describing the process are introduced and the derived algorithm is used to interpret an IBIC experiment aimed to characterise a 4H-SiC diode array.

**Experimental**

The sample under test consists in an array of 16 Schottky electrodes with total area 0.4×0.4 mm$^2$ fabricated by Alenia Marconi on an n-type (nominal donor concentration: $N_D$=5·10$^{14}$ cm$^{-3}$, 30 µm thick) epitaxial layer produced by CREE Research company. A picture of the device is reported in Fig. 1a; details on detector fabrication can be found in [6]. The guard ring is 60 µm wide and separated from the electrodes by a 20 µm gap. The back Ti-Pt-Au ohmic contact is deposited on the C face of the n+ substrate ($N_D \approx 6.8 \cdot 10^{18}$ atoms cm$^{-3}$); the Au front Schottky contact is 100 nm thick.

The electrical characterization of the device is described in details in [7]: the rectifying junctions exhibit 1.2 eV Schottky barrier height and reverse current density lower than 1 nA cm$^{-2}$ at 100 V bias voltage.

The measurements were carried out at the ion microbeam facility of the Ruđer Bošković Institute in Zagreb (HR) using a 1.5 MeV proton beam focused to a spot size of less than 3 µm and raster scanned onto the device surface. The ion currents were maintained below 0.1 fA in order to avoid pileup effects and the measurements were performed at room temperature in high vacuum conditions (pressure below 10$^{-3}$ Pa).

The charge collection efficiency was evaluated by comparing the IBIC signal from the 4H-SiC diode with the signal from a silicon surface barrier detector, for which a complete charge collection is assumed, taking into account the different energy to create electron/hole pairs in Si (3.6 eV) and in 4H-SiC (7.78 eV) and the energy lost in the entrance window, following the procedure described in detail in [8]. In this paper we report on the measurements carried out on two diodes of the 4×4 array connected to two charge sensitive electronic chains.

FEMLAB software package was used to solve by the finite element methods the coupled Poisson's and continuity equations, the relevant weighted potential distribution and the adjoint equations, and to evaluate the charge collection efficiency maps [9]. Physical input parameters were extracted from [10]; the generation profile was evaluated from SRIM Monte Carlo simulation [11].

**Results and discussion**

**a) Theory**

Since the charge induced at the sensing electrode is due to the motion of electron/hole pairs in a medium permeated by a static electric field, an interpretative theoretical model of IBIC experiments has to be founded on the basic laws of electrostatics, under the following assumptions:

1. the perturbation of the electromagnetic field within the semiconductor due to the excess charge produced by ionization is negligible

2. the electromagnetic field propagates instantaneously in the device

3. the electrodes are connected to ideal voltage sources of dc potential, so that the motion of charge carriers does not affect the electrode potentials.

The Shockley-Ramo's theorem [5], which is essentially a lemma of the Green's reciprocity theorem, has been extensively used to evaluate charge collection in fully depleted devices. More recently, the Gunn's theorem has been brought to light [12], allowing a rigorous interpretation of IBIC and time resolved IBIC (TRIBIC) experiments in partially depleted diodes [13][14] and in MOS devices [15]. However, all these papers concern one-dimensional problems; if two-dimensional IBIC experiments are considered, the "weighting field" concept can suitably provide a mathematical formalism to evaluate the spatial distribution of the induced charge.

To illustrate this method, let us consider an example which follows closely the example discussed by V. Radeka [16]. It refers to a planar n-type Schottky diode with two virtually grounded sensing electrodes ($S_1$ and $S_2$), a grounded guard ring (G), and a back electrode (B) at a bias voltage V. Fig. 1b schematically shows the system geometry with the map of the actual potential $\phi$.

The basic expression resulting from the Gunn's theorem [12]:

$$(1) \quad I_{S_1}(t) = -q \cdot \left. \frac{\partial \overline{E}}{\partial V} \right|_{V_B=V, V_{G,S2}=0} \cdot \frac{d\overline{r}}{dt}$$

allows the calculation of the current $I_{S1}$ induced at the sensing electrode $S_1$ due to the motion of a point charge q along a line segment $d\overline{r}$ with velocity $v = \frac{d\overline{r}}{dt}$.

$\left. \frac{\partial \overline{E}}{\partial V} \right|_{V_B=V, V_{G,S2}=0}$ is the "Gunn's weighting field" (expressed in units of cm$^{-1}$), and it is defined as the derivative of the electric field **E** with respect to the voltage at the sensing electrode, assuming the potentials ($V_{S2}$, $V_G$ and $V_B$) of the other conductors as fixed.

The charge $Q_{S1}$ induced at the sensing electrode by the motion of a charge q from the initial position **r$_i$** to the final position **r$_f$** is obtained from eq. (1), by integrating in time the induced current between the initial ($t_i$) and final ($t_f$) times:

$$Q_{S1}(r_f - r_i) = \int_{t_i}^{t_f} I(t)dt = -q \cdot \int_{t_i}^{t_f} \left.\frac{\partial \overline{E}}{\partial V}\right|_{V_B=V,V_{G,S2}=0} \cdot \frac{d\overline{r}}{dt}dt = -q \cdot \int_{r_i}^{r_f} \left.\frac{\partial \overline{E}}{\partial V}\right|_{V_B=V,V_{G,S2}=0} d\overline{r}$$

(2)

$$= -q \cdot \left[ \left.\frac{\partial \phi}{\partial V}\right|_{r_f} - \left.\frac{\partial \phi}{\partial V}\right|_{r_i} \right]$$

The induced charge is then simply given by the difference in the Gunn's weighting potential $\frac{\partial \phi}{\partial V}$ between the initial and final position of the moving charge. As a consequence, the weighing potential mapping allows the evaluation of Q for any charge generation point in the device. Fig. 2a,b show the weighted potential distribution derived from the electrical potential mapped in Fig. 1b, when the two sensing electrodes $S_1$ and $S_2$ are considered separately.

Let us consider an electron/hole pair generated in position A underneath the sensing electrode $S_1$. The trajectory of the hole is upward along line 1 in Fig. 1b and Fig. 2a; it terminates at the sensing electrode, i.e. $r_{fAh} = (x = x_A, y = d)$, where $\frac{\partial \phi}{\partial V} = 1$. As a consequence, the charge induced by the hole $Q_{S1}^h$ is given by:

(3) $$Q_{S1}^h(r_{fAh} - r_{iA}) = -q \cdot \left[ \left.\frac{\partial \phi}{\partial V}\right|_{r_{fA}} - \left.\frac{\partial \phi}{\partial V}\right|_{r_{iA}} \right] = -q \cdot \left[ 1 - \left.\frac{\partial \phi}{\partial V}\right|_{r_{iA}} \right].$$

If an electron is generated at position A, its downward drift trajectory ends in the neutral region, where the weighted potential is zero. In this case, the induced charge $Q_{S1}^e$ is:

(4) $$Q_{S1}^e(r_{fAe} - r_{iA}) = q \cdot \left[ \left.\frac{\partial \phi}{\partial V}\right|_{r_{fAe}} - \left.\frac{\partial \phi}{\partial V}\right|_{r_{iA}} \right] = q \cdot \left[ 0 - \left.\frac{\partial \phi}{\partial V}\right|_{r_{iA}} \right]$$

It follows that the electron-hole motion within the detector induces a total charge $Q_{S1} = Q_{S1}^h + Q_{S1}^e = -q$ (i.e. total charge collection occurs) only at the sensing electrode $S_1$, while no signals is detected at $S_2$. Vice versa, a complete charge collection occurs at electrode $S_2$ with no signal at $S_1$ if the electron/hole pair is generated at position C (see Figs. 1b and 2b).

Finally, no signals can be detected if the generation occurs at position F, i.e. underneath the guard electrode, since the trajectories of the carriers cross regions where the weighted potentials relevant to both electrodes are null.

As expected, for charges generated in the gaps between G and $S_{1,2}$, the total induced charge assumes values comprised between 0 and q.

If a frontal IBIC experiment is performed on the device illustrated in the previous example, assuming negligible plasma and/or high injection effects, the contribution of electron-hole pairs generated by ionisation can be integrated over the ionization profile and the total charge induced at the sensing electrode can be written as follows:

$$(5) \quad Q(x) = -q \cdot \int_0^{\Gamma} B(x,y) \left[ 1 - \left. \frac{\partial \phi}{\partial V} \right|_{x,y} \right] dy + q \cdot \int_0^{W(x)} B(x,y) \left[ \left. \frac{\partial \phi}{\partial V} \right|_{x,y} \right] dy$$

where $B(x,y)$ is the ionisation profile (along the y direction) generated by an incident ion hitting the device at position x, and $\Gamma$ is the ion penetration depth. $W(x)$ is the depth where the weighted potential is not null (which roughly corresponds to the depletion layer). Equation (5) holds only if $\Gamma < W(x)$.
8

If the generation occurs in regions between the sensing electrodes and the guard ring, or in the case of ionization profiles extending beyond the depletion layer, holes generated in the diffusion region can reach the edge of the depletion layer by diffusion, and hence contribute to the induced charge. This is a well known effect, already considered in the pioneering paper of M. Breese [17], and extensively used to evaluate the diffusion lengths of minority carriers in partially depleted detectors [18].

As a consequence, in order to have a complete model of the charge induction mechanism in a partially depleted device, the Gunn's equation has to be coupled with the continuity equations of semiconductor free carriers in the quasi-steady-state mode, as effectively described by Prettyman [19] and applied to interpret IBIC/TRIBIC experiments in pristine or damaged devices [13], [20].

In summary, considering the above mentioned assumptions, the continuity equations for electrons and holes can be decoupled from Poisson's equation and, under the hypothesis that the generation-recombination terms can be linearized, the adjoint continuity equations can be constructed.

If the adjoint generation terms involve the Gunn's electric field, it is possible to demonstrate that the Green's functions (namely, $n^+$ and $p^+$ for electrons and holes, respectively) of the continuity equations represent the charge pulse at time t, induced at the sensing electrode by the motion of electrons and holes generated in **r** at t=0. In other words, charge collection efficiency mapping can be effectively carried out by solving two time dependent linear differential equations [19][13].

The steady state solutions of these equations satisfy the following time independent linear differential equations:

(6)
$$\begin{cases} -\mathbf{v}_n \cdot \nabla n^+ + \nabla \cdot (D_n \nabla n^+) - \dfrac{n^+}{\tau_n} + G_n^+ = 0 \\ +\mathbf{v}_p \cdot \nabla p^+ + \nabla \cdot (D_p \nabla p^+) - \dfrac{p^+}{\tau_p} + G_p^+ = 0 \end{cases}$$

$$\begin{cases} -\mathbf{v}_n \cdot \nabla n^+ + \nabla \cdot (D_n \nabla n^+) - \dfrac{n^+}{\tau_n} + G_n^+ = 0 \\ +\mathbf{v}_p \cdot \nabla p^+ + \nabla \cdot (D_p \nabla p^+) - \dfrac{p^+}{\tau_p} + G_p^+ = 0 \end{cases}$$

where the source terms are:

(7)
$$\begin{cases} G_n^+ = +\mathbf{v}_n \cdot \dfrac{\partial \mathbf{E}}{\partial V} - \nabla \cdot \left[ D_n \cdot \dfrac{\partial \mathbf{E}}{\partial V} \right] \\ G_p^+ = +\mathbf{v}_p \cdot \dfrac{\partial \mathbf{E}}{\partial V} + \nabla \cdot \left[ D_p \cdot \dfrac{\partial \mathbf{E}}{\partial V} \right] \end{cases}$$

$\tau_{n,p}$ are the carrier lifetimes, $v_{n,p} = \mu_{n,p} \cdot E$ are the drift velocities, $\mu_{n,p}$ the mobilities and $D_{n,p}$ the diffusivities for electrons and holes, respectively. The boundary conditions of above listed equations must be homogeneous and Dirichlet-type for the electrodes and at the inter-electrode gap, and symmetric and Neumann-type for the left-right extremes.

Figs. 3a,b show the Charge Collection Efficiency (CCE) map of the device at a bias voltage of V=100 V, considering the two sensing electrodes separately. A comparison of these figures with Fig. 1 and Figs. 2a,b highlights the different extension of the depletion region with respect to the active region (i.e. the region

whereCCE>0), due to the diffusion of minority carriers generated into the neutral region towards the depletion region.

In order to clarify this point, let us consider the CCE profiles along line 1 for electrons and holes at different bias voltages (Figs. 4a,b,c), relevant to the sensing electrode $S_1$.

Since the carrier lifetime (>100 ns [14]) is much longer than the drift time (<1 ns), nearly all the free carriers generated into the depletion region are collected. This is the origin of the plateaux extending all through the extension of the depletion region. Such plateaux increase their width as the bias voltage increases.

Electrons generated within the neutral region cannot enter into the depletion region because of the opposite electrical force; as a consequence, they cannot induce charge at the sensing electrode. On the contrary, holes generated in the neutral region can diffuse to the active region, i.e. where the weighted potential is not zero, and drift towards the sensing electrode, inducing a signal. Therefore, the exponential tail of the hole contribution is determined by the carrier diffusion length as demonstrated in previous publications [18].

Finally, in the case of frontal experiments, a complete interpretation of IBIC profiles along the lateral direction can be obtained using the following generalized equation (5):

(8) $\quad Q(x) = -q \cdot \int_0^\Gamma B(x,y) \cdot [p^+(x,y) + n^+(x,y)] dy$.

**b) Experimental results**

Fig. 5 shows the experimental IBIC maps at 100 V bias voltage from the region highlighted in Fig. 1, relevant to two Schottky diodes. The detectors show a uniform response, apart "shadowing effects" of the gold wire used for bonding.

Fig. 6 shows the experimental and theoretical CCE vs. the applied bias voltage of detector 1, obtained by scanning the proton beam onto area R highlighted in Fig. 5. The signal saturation is achieved at a bias voltage of 180 V, as expected when the active region extends beyond the ionization profile (see Fig. 4d). At lower bias voltages, a partial charge collection occurs for charges generated in the neutral region. A suitable fit of this curve can be carried out either using the commonly used drift-diffusion model or iterating the algorithm defined by eqns. (6) and (7), and setting the carrier lifetime as the fitting free parameter. Both methods converge to a fitting curve corresponding to a value of the hole diffusion length of 8.7 $\mu$m. The corresponding hole lifetime value ($\tau_p$ = 250 ns) was then used to evaluate the CCE profile shown in Figs. 3a,b.

Finally, Fig. 7 shows the CCE profile at V=100 V from a scan of the proton beam along the line perpendicular to the sensing electrodes highlighted in Fig. 5. The theoretical CCE profile obtained from eq. (8) fits the experimental data very satisfactorily. These results validate the mathematical model and allow the extension of the active region to be accurately defined.

## Conclusions

The algorithm presented in this paper enables the efficient interpretation and modeling of IBIC experiments. It is based on the solution of two uncoupled, linear and time independent differential equations, provided that the electrostatic field and the "Gunn's weighting field" within the device are known. The method is efficient to map the charge collection efficiency in semiconductors and, since it involves linear equations, it is potentially suitable to solve problems with complex geometries with reduced computational effort. The mapping method has proven effective to analyze the benchmark IBIC experiment described in this paper, allowing the measurement of minority carrier diffusion lengths in 4H-SiC Schottky diodes and the accurate definition of the active region underneath the diode array.

**Figure Captions**

Fig. 1  (a) Photograph of the diode array; the red rectangle indicates the investigated region (see Fig. 5). (b) Cross section scheme of the device with the relevant electrical connections. The map represents the electric potential map when the applied bias voltage is V=100 V.

Fig. 2  Map of the weighted potential when the applied bias voltage is V=100 V. The sensing electrodes are labelled $S_1$ (a) and $S_2$ (b).

Fig. 3  Theoretical CCE map for an applied bias voltage of 100 V. The sensing electrodes are labelled $S_1$ (a) and $S_2$ (b).

Fig. 4  (a,b,c) CCE profiles along line 1 at different bias voltages: p=hole contribution, n=electron contribution. (d) Bragg curve of 1.5 MeV protons in 4H-SiC.

Fig. 5  Experimental frontal CCE maps relevant to sensing electrodes S1 and S2 when the applied bias voltage is 100 V. Region R indicates the area irradiated to evaluate data in Fig. 6. The white rectangle indicate the scanned region to evaluate the CCE profile shown in Fig. 7.

Fig. 6  Experimental (marks) and theoretical (continuous line) CCE's at different bias voltages collected at electrode $S_1$, relevant to region R in Fig. 5.

Fig. 7  Experimental (marks) and theoretical (continuous line) CCE profiles along the white line perpendicular to the electrodes shown in Fig. 5.

**Figures**

**Fig. 1a**

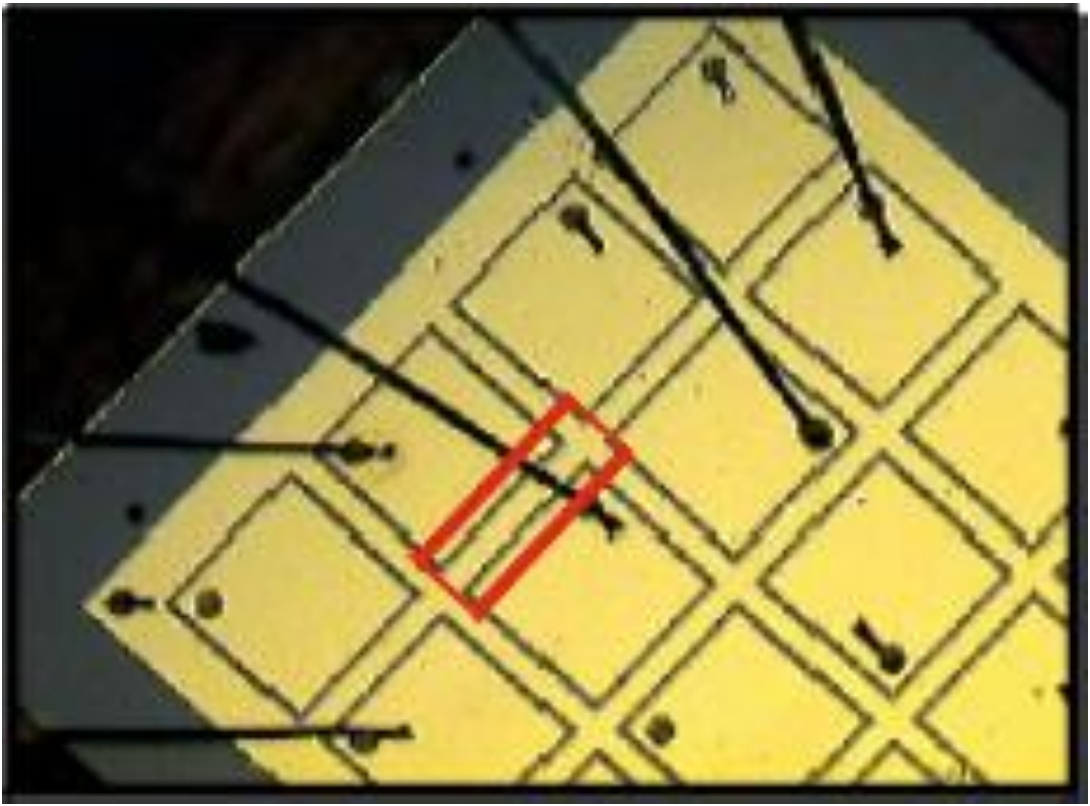

**Fig. 1b**

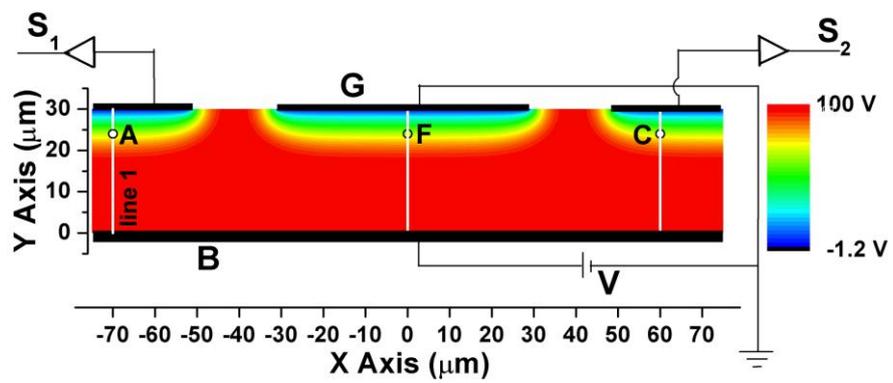

**Fig. 2a**

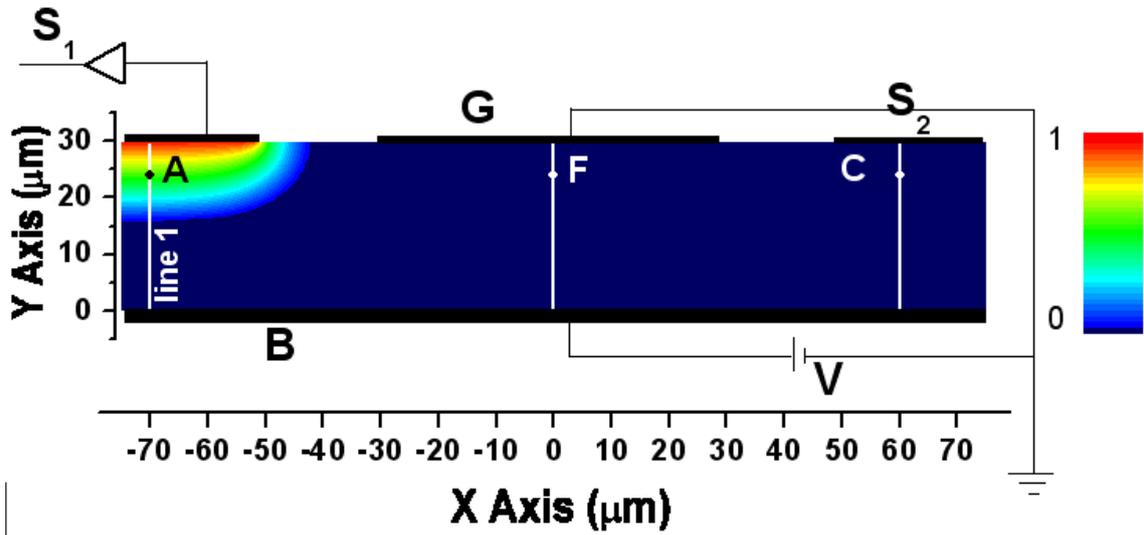

**Fig. 2b**

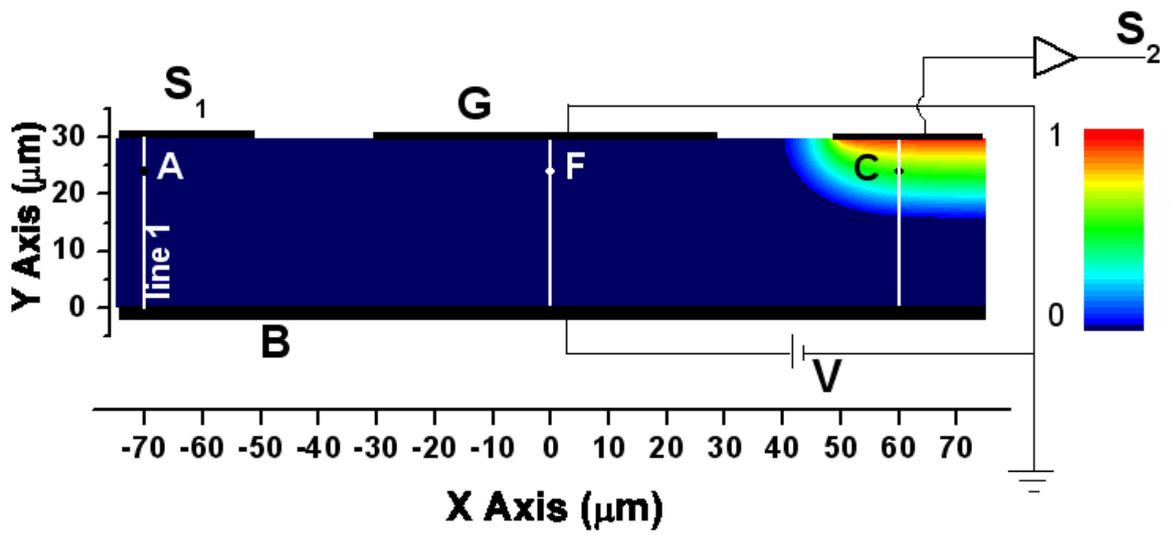

**Fig. 3a**

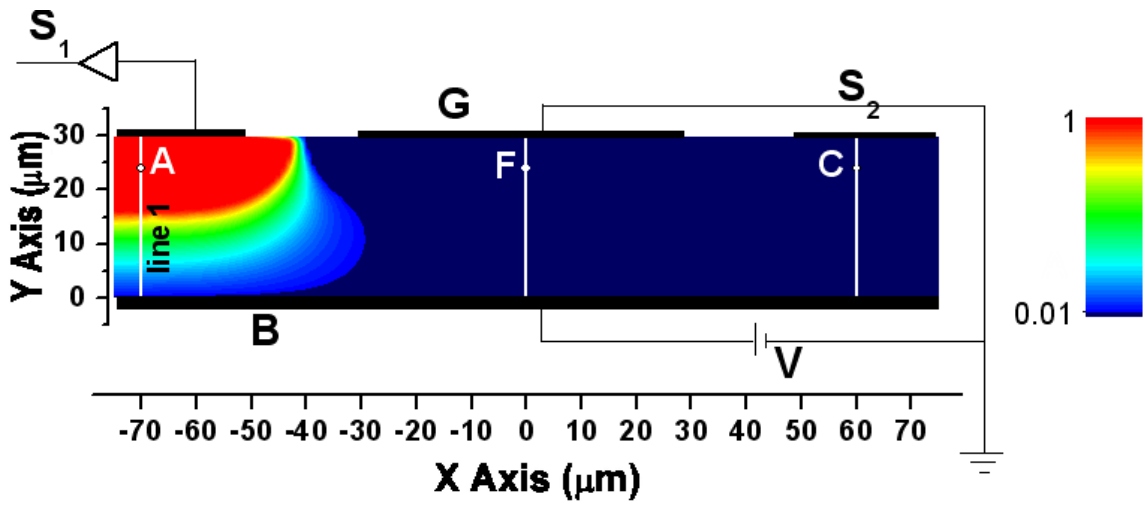

**Fig. 3b**

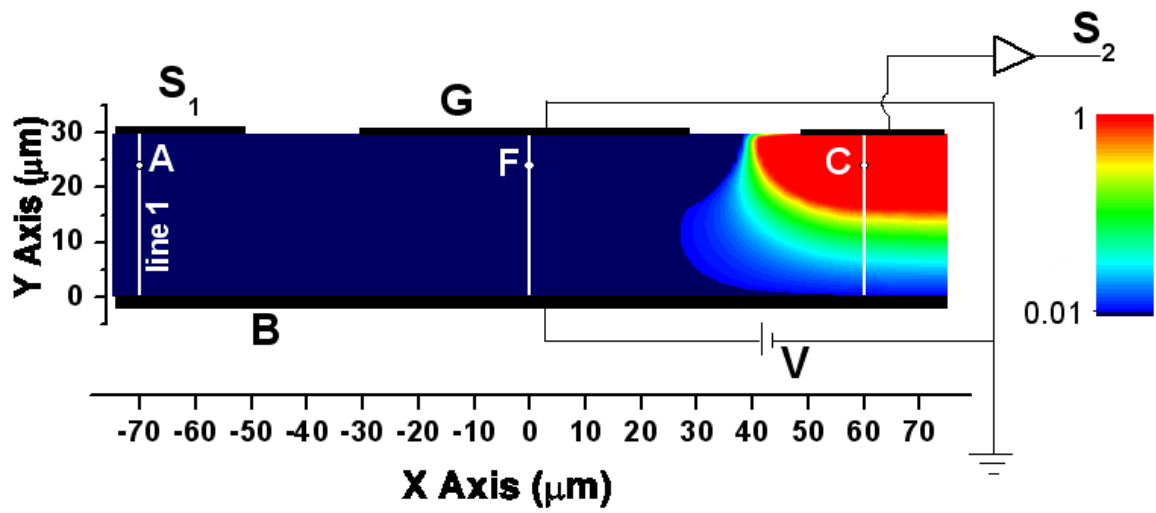

**Fig. 4**

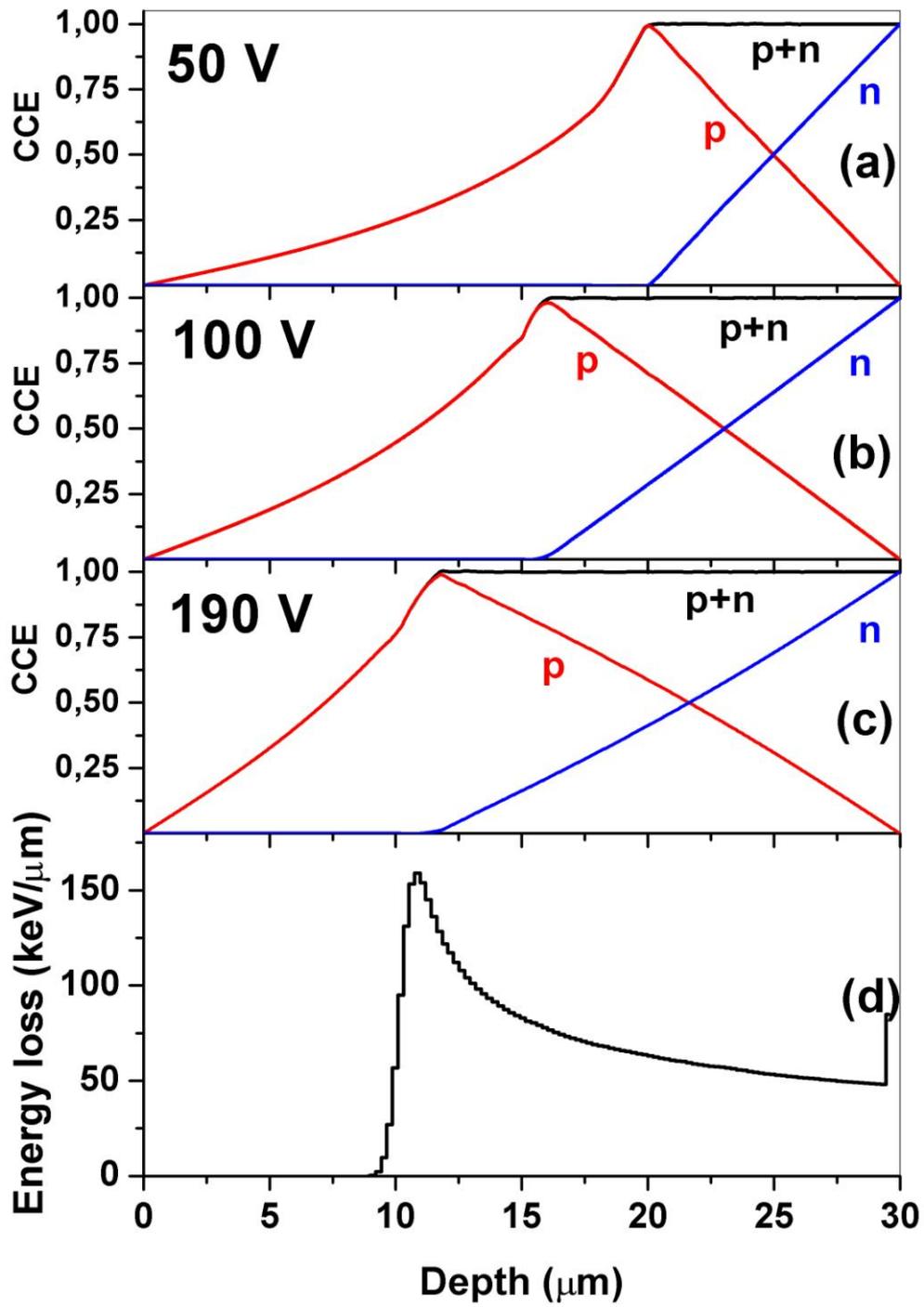

**Fig. 5**

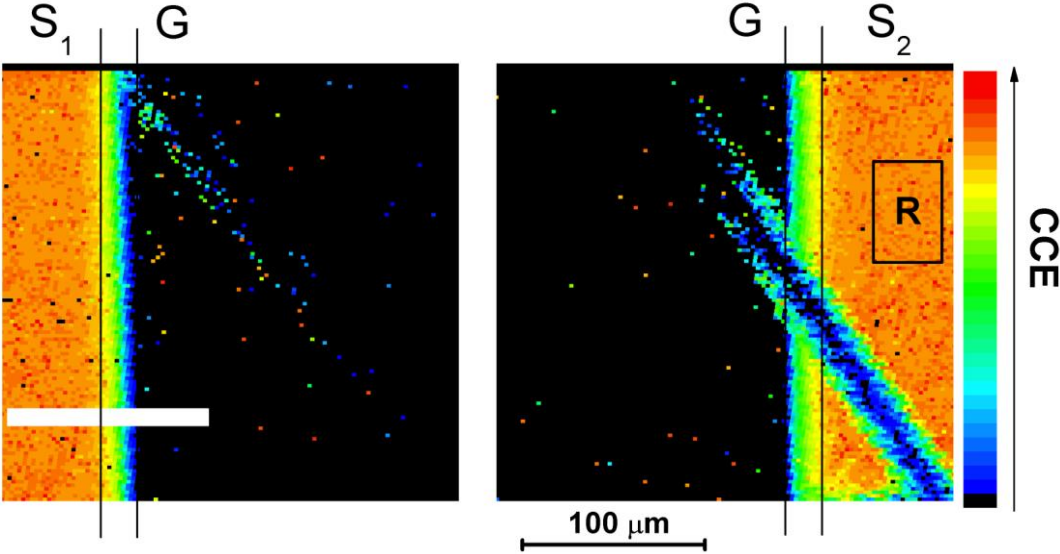

**Fig. 6**

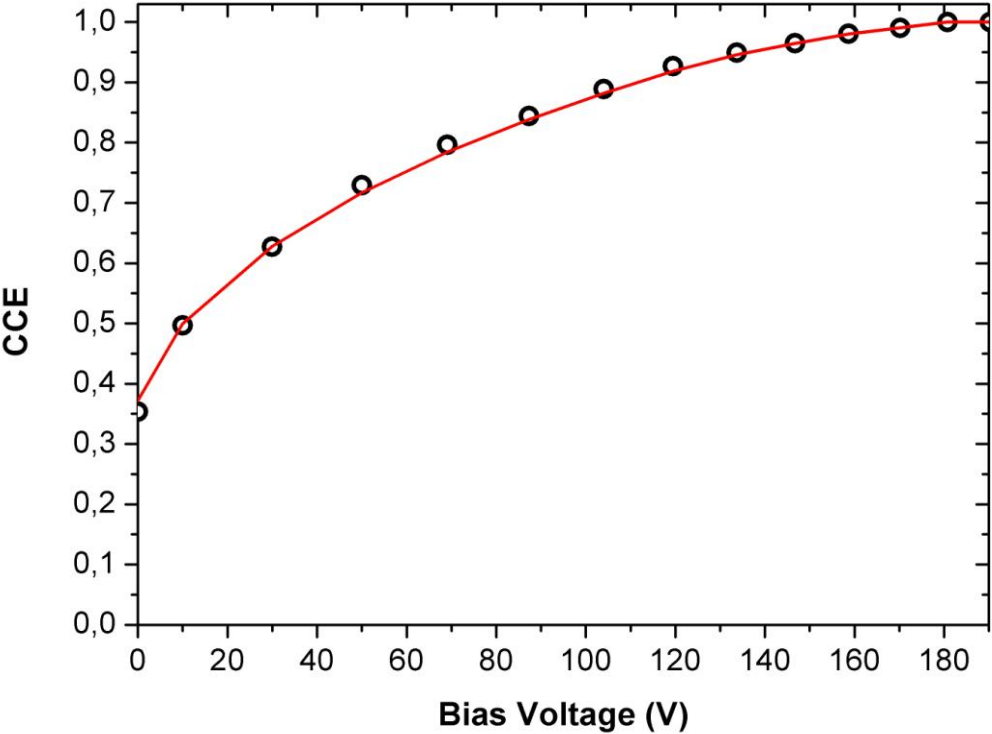

**Fig. 7**

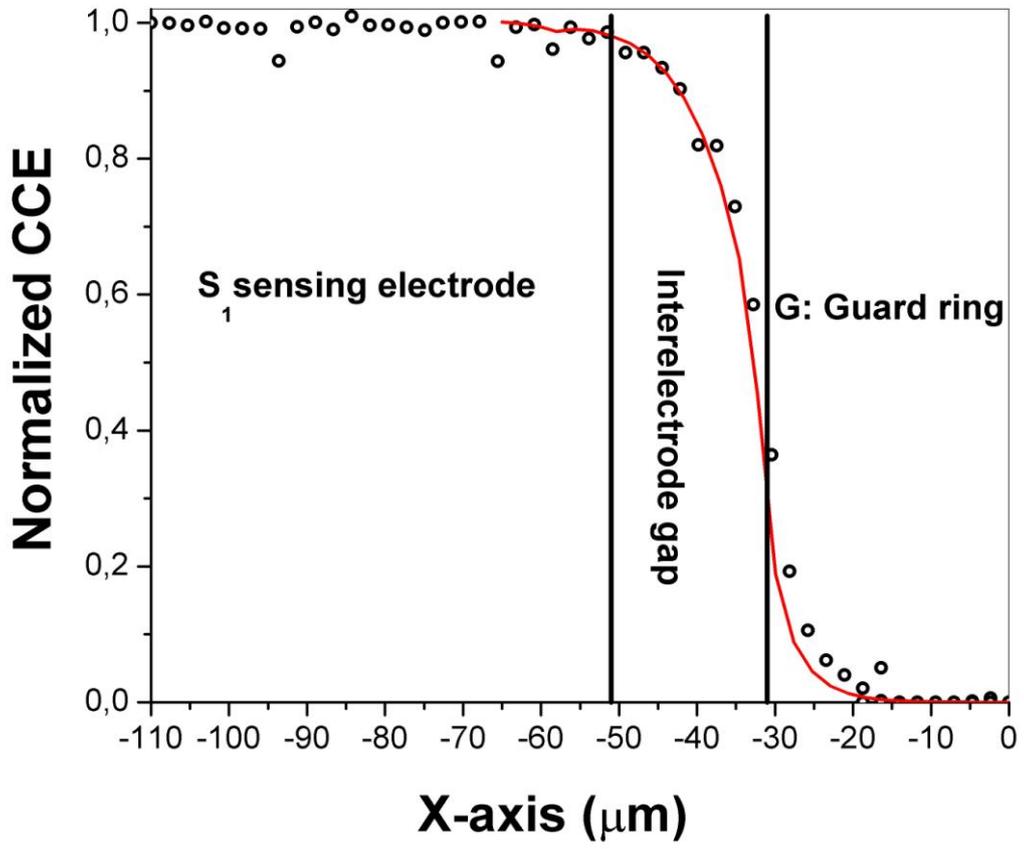